\documentstyle {slides}
\begin{document}
\newcommand{\be}{\begin{equation}}
\newcommand{\ee}{\end{equation}} 
\hfill        ICHEP06, 27.07.2006

\begin{center}
{\large {\bf Cherenkov gluons as a probe of nuclear matter}}\\
      (hep-ph/0507167; 0602135)

 I.M. Dremin
\end{center}

Experimental data (early days + RHIC).\\
1. Stratospheric event at 10$^{16}$ eV (1979). Its wavelet image (1997).
(LHC - 2008!)  \\
2. Pseudorapidity distribution of dense isolated groups in pp at 360 GeV 
(1989).\\
3. Wavelet image of the PbPb event at 158 GeV shows ring-like structure (2001).   \\
4. (Fig.) Two-bump distribution at RHIC (central AuAu at 20 TeV - $c_w=0.33v$) 
(2004).

{\bf Theory}\\
If $z$-axis is along the body propagation, then emission 
in an {\bf infinite} medium {\bf at rest} 
is directed along the cone with the polar angle $\theta $:
\be
\cos \theta = \frac{c_w}{v}.     
\ee
\newpage 
$v$ is the body velocity.
For Cherenkov photons $c_w=c/n$, $n$ is the index of refraction. 
\be
n(\omega )=1+\Delta n =1+\frac {2\pi N}{\omega ^2}F(\omega ).   \label{nom}
\ee
$\omega $ is the photon frequency, $N$ is the density of the scatterers
(inhomogeneities) of the medium. The forward scattering amplitude $F(\omega )$ 
is normalized as
\be
{\rm Im} F(\omega )= \frac {\omega }{4\pi }\sigma (\omega ),    \label{opt}
\ee
where $\sigma (\omega )$ is the total cross section of photon interaction
in the medium.

{\bf For hadrons}
\be
{\rm Re} n(\omega )=1+\Delta n_R(\omega )=1+\frac {3m_{\pi }^3}{8\pi \omega}
\sigma (\omega )\rho (\omega )\nu ,     \label{ren}
\ee
where $\rho (\omega )={\rm Re} F/{\rm Im }F$ and now $F(\omega ) \;(\sigma (\omega ))$ 
is the pion-nucleon forward amplitude (cross section), $\nu $-the number of 
scatterers within a nucleon.  

The resonance region:
\be
\Delta n_R^r=\frac {3m_{\pi }^3}{2\omega _r^2\Gamma}. \label{del}
\ee

\newpage
RHIC - Low energy gluons.

 For a single resonance at energy $E_R$ and with width $\Gamma _R$ one gets 

${\rm Re} n(E)=1+$
\be
\frac {2J+1}{(2s_1+1)(2s_2+1)}\cdot \frac {3\mu ^3\Gamma _R\nu }
{E_R^2}\cdot \frac{E-E_R}{E[(E-E_R)^2+\Gamma _R^2/4]}.   
\ee
Here $J$ is the spin of the resonance, $s_i$ are the spins of "incident 
particles". We have replaced $N$ by the number of partonic scatterers $\nu $
within a single nucleon volume $4\pi /3\mu ^3$ with $\mu $ the pion mass.

${\rm Re} n>1$ for $E>E_R$ only. It is 1 at $E=E_R$, increases to the maximum 
approximately at the resonance half-width
\be 
E_0=E_R+\frac {\Gamma _R}{2}(1-\frac {\Gamma _R}{2E_R})    
\ee
and decreases as $O(1/E^2)$. For $\Gamma _R\ll E_R$ its value at the maximum is
\be
{\rm Re} n(E_0)\approx 1+\frac {2J+1}{(2s_1+1)(2s_2+1)}\cdot \frac {3\mu ^3\nu }
{E_R^3}.                     
\ee
The average value of ${\rm Re} n$ is smaller than its value at the maximum. 
Therefore, from experimental estimate of this average value equal to 3 one 
gets the lower bound for the effective number of partons within the nucleon 
volume $\nu $
\be
\nu >\frac {(2s_1+1)(2s_2+1)}{2J+1}\cdot \frac {2E_R^3}{3\mu ^3}.  
\ee
$\nu > 40$ for $\rho $, $>20$ for all resonances included.

The average value of ${\rm Re} n$ is
\be
\langle {\rm Re} n\rangle \approx 1+\frac {2J+1}{(2s_1+1)(2s_2+1)}\cdot \frac 
{6\mu ^3 }{\pi E_R^3}\langle \nu \rangle.    
\ee
$\langle \nu \rangle=60$ for $\rho $, $=30$ for all.

Another information can be obtained from the height of the peaks in STAR Fig..
For narrow rings ($\delta \ll r_i$) the height of the maximum over the
minimum is easily determined as
\be
h_{max}=\sqrt {2r_1\delta }-\delta .    
\ee
With $h_{max} \approx 1.6 - 1.2 = 0.4$ and $r_1 = 1.2$ in STAR Fig. one obtains
the width:
\be
\delta \approx 0.1  \;\;\;\; (\ll r_1).        
\ee

Dispersion
\be
\delta _d=\int _0^{\delta _d}d\theta =\cot \theta _c\int _0^{\infty }
\frac {1}{n} \frac{dn}{dE} dE.      
\ee
If the Breit-Wigner expression for $n(E)$ is used, the result is
\be
\delta _d=0.   
\ee

The finite free path length for partons
\be
\delta _f \sim \frac {\lambda }{R_f}.    
\ee
For $\lambda \sim 2/E_R$ and $\delta _f < 0.1$ one gets the estimate for
the free path length
\be
R_f>20/E_R \sim 4.5/\mu \sim 7\cdot 10^{-13} cm.   
\ee
Finally, the width of the ring can become larger due to resonance decays.

The energy loss can be calculated using the standard formula
\be
\frac {dE}{dx}=4\pi \alpha _S\int _{E_R}^{E_R+\Gamma _R}E\left (1-\frac {1}
{n^2(E)}\right )dE\approx 1 GeV/fm.
\ee

\newpage

\begin{center}
{\bf CONCLUSIONS}
\end{center}

QM properties  revealed by Cherenkov gluons:

1. The nuclear index of refraction $n\approx 3$.
Determined by the $\Delta y$-distance between the peaks (ring diameter).

2. The density of partons $\nu \approx 30$ (per nucleon).
Use $n$ (with resonance amplitudes).

3. The energy loss $dE/dx \approx 1$ GeV/fm.
Determined by the standard formula with $n$.

4. The free path length for partons $R_f \sim 7 $ fm.
The height of the peaks (ring width).

5. Ring content - "shifted" resonances.

What is missing?

1. The two-dimensional data in the plane perpendicular to
the associated jet axis.

2. Event-by-event correlation analysis within the ring.

3. Search for resonance content in the ring.

{\bf References} added to the talk slides:

1. I.M. Dremin, Pis'ma v ZhETF 30 (1979) 152; JETP. Lett. 30 (1979) 140.

2. I.M. Dremin, Yad. Fiz. 33 (1981) 1357; Sov. J. Nucl. Phys. 33 (1981) 726.

3. I.M. Dremin, Nucl. Phys. A 767 (2006) 233.

\end{document}